\documentclass[11pt,aps]{revtex4}
\usepackage{graphicx}

\def\b{\beta}

\def\D{\Delta}
\def\e{\epsilon}

\def\g{\gamma}
\def\G{\Gamma}

\def\l{\lambda}

\def\m{\mu}
\def\n{\nu}
\def\r{\rho}
\def\s{\sigma}

\def\th{\theta}

\def\D{\Delta}
\def\G{\Gamma}

\def\GG{{\rm I}\!\G}

\newcommand{\fr}{\frac}
\newcommand{\bc}{\begin{center}}
\newcommand{\ec}{\end{center}}
\newcommand{\be}{\begin{equation}}
\newcommand{\ee}{\end{equation}}
\newcommand{\bea}{\begin{eqnarray}}
\newcommand{\eea}{\end{eqnarray}}
\newcommand{\bei}{\begin{itemize}}
\newcommand{\eei}{\end{itemize}}
\newcommand{\been}{\begin{enumerate}}
\newcommand{\een}{\end{enumerate}}
\newcommand{\no}{\nonumber}
\newcommand{\CK}{\mathcal{K}}
\newcommand{\CP}{\mathcal{P}}

\def\slash#1{#1\!\!\!/\!\,}
\newcommand{\intk}{\int\frac{d^3k}{(2\pi)^3}}

\begin{document}

\title{Effect of noncommutative Chern-Simons term on the magnetic moment of
planar fermions}
\author{T. Shreecharan}

\email{shreet@prl.res.in}

\affiliation{Physical Research Laboratory, Navrangpura, 
Ahmedabad, India 380 009}
 
\begin{abstract}
We study the effect of noncommutative Chern-Simons term on 
fundamental fermions. In particular the one-loop contribution to 
the magnetic moment to order $\th$ is calculated. 
 \keywords{Noncommutative; Chern-Simons; Magnetic Moment.}
\end{abstract}

 \maketitle


\section{Introduction}

Noncommutative (NC) field theories are defined on a space-time 
characterized by the commutation relation: $[x^\m,x^\n]=i 
\th^{\m\n}$; here the noncommutative parameter $\th^{\m\n}$ is 
treated as a constant. Quantum field theories defined on such a 
space-time show many interesting features that are absent in 
ordinary field theories and have recently been vigorously pursued 
\cite{ncrevs}. Though the idea itself is old \cite{Snyder}, the 
recent impetus in this field can be traced to its emergence in 
string/M(atrix) theory \cite{Seiberg,Schwarz}. Keeping the string 
theoretical considerations aside, NC field theories themselves 
pose interesting challenges and features. The most discussed 
properties happen to be the loss of unitarity if one includes 
time also to be noncommuting \cite{Mehen}. Here it must be 
pointed out that, the reported violation of unitarity can be 
overcome if the quantization conditions are defined in a proper 
manner \cite{Fredenhagen}. Furthermore, contrary to the initial 
motivation for NC theories \cite{Snyder}, namely regulating the 
ultraviolet divergences; there appear new divergences compared to 
the corresponding commutative theory. This is due to the coupling 
of the ultraviolet sector to the infrared, dubbed UV/IR mixing 
\cite{Minwalla}.

In the present work we are concerned with NC Chern-Simons (CS) 
term coupled to planar fermions in the fundamental 
representation. It is well known that a CS term can be added to 
the Maxwell lagrangian in 2+1 dimensions which can provide mass 
to the gauge field independent of Higgs mechanism 
\cite{Deser,Hagen}. Apart from this CS field theories have found 
applications in various branches of mathematics and physics, 
notably in knot theory \cite{Witten} and as an effective 
description of quantum Hall effect (QHE) \cite{Dunne}. The 
particular use of NCCS theory in the description of QHE, as shown 
in \cite{Susskind} and further pursued in \cite{Nair} is quite 
interesting. 

Independent of its usage in the description of QHE, a number of 
formal perturbative studies have been conducted 
\cite{Bichl,Grandi,Kaminsky}. Here we calculate the MM of 
fermions coupled to the NCCS gauge field. The corresponding 
calculation in the commutative non-Abelian and the Abelian cases 
was carried out in \cite{Chaichian1} and \cite{Chaichian2,Rashmi} 
respectively. 

The manuscript is organized as follows. In the next section we 
set up the NCCS action and state the Feynman rules. In section 
III the vertex diagram is evaluated. Our conclusions and 
discussion of the results are presented in section IV.

\section{The noncommutative action and Feynman rules}

\begin{figure}
\begin{tabular}{cc}
\begin{minipage}{2in}
\centering
\includegraphics[width=1.5in,height=1.5in]{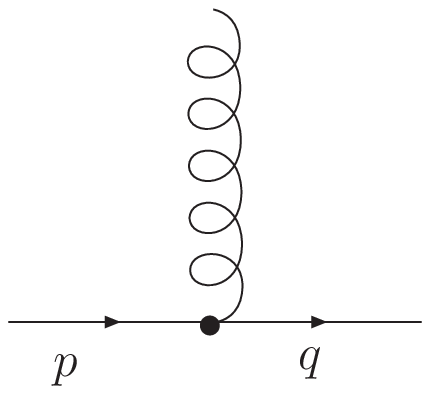}
\caption{The fermion-photon vertex.} \label{fpv}
\end{minipage}
& {\large $\equiv i e \g_\m \exp\left[\frac{i}{2}p \times q
\right]$.} \\
\begin{minipage}{2in}
\centering
\includegraphics[width=2in,height=2in]{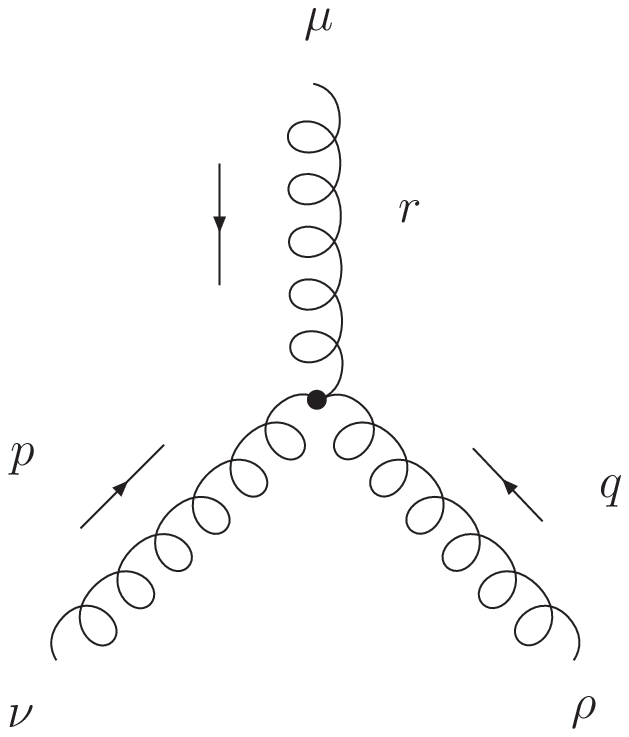}
\caption{The three gauge boson vertex.} \label{sf3}
\end{minipage}
& {\large $\equiv 2 i e M \e^{\n\m\r} \sin\left[(p \times 
r)/2\right]$.}
\end{tabular}
\end{figure}

The NC action for a pure CS theory in Minkowski space reads
 \be
S_{CS}= \frac{M}{2} \int d^3x \, \e^{\m\n\r} \, \left[ 
A_{\m}\star\partial_{\n}A_{\r} + \frac{2ie}{3} A_{\m} \star 
A_{\n} \star A_{\r} \right].
 \ee
Here $M$ is the CS coefficient and $e$ the coupling constant. 
Note that due to the noncommuting nature of the fields there 
emerges a three gauge boson interaction term, akin to the 
commutative non-Abelian theory. Let us consider the three photon 
interaction term. Since we deal with only spatial 
noncommutativity one of the star product can be dropped. The rest 
of the fields being bilinear gives usual product therefore three 
gauge boson vertex can be dropped \cite{apb}. This has to be 
contrasted with the NCCS actions based on Seiberg-Witten map that 
also leads to the extinction of the three photon vertex 
\cite{Grandi}. This equivalence of the pure NCCS theory has been 
shown to persist at the quantum level as well \cite{Kaminsky}. To 
proceed further we add a gauge fixing term
 \be
S_{gf} = -\frac{1}{2\xi} \int d^3x \, (\partial_\m A^\m \star
\partial_\n A^\n).
 \ee
The Dirac action gives the matter content of the theory
    \be \label{nccsspinor}
S_{Dirac} = \int d^3x \, \bar{\psi}\star(i \g^\m D_\m - m)\psi ,
    \ee
where $D_\m \psi=\partial_\m\psi - i e A_\m \star\psi$. The gauge 
($G_{\m\n}(p)$) and Fermionic ($S(p)$) propagators are 
 \be
    iG_{\m\n}(p)= - \frac{1}{M}\e_{\m\n\r}\frac{p^{\r}}{p^2 + i \e}
 \ee
and 
 \be
iS(p)=\fr{i(\slash{p}+m)}{p^2-m^2+i\e}, 
 \ee 
respectively. The gauge propagator is defined in the Landau gauge 
($\xi=0$) since it was shown that other gauges can introduce 
spurious infrared divergences \cite{Pisarski}. The interaction 
vertices are depicted in Figs. (\ref{fpv}) and (\ref{sf3}). In 
what follows the $i \e$ term in the propagators is to be 
understood.

\section{The Abelian vertex}

\begin{figure}
\begin{center}
\includegraphics[height=3.5in,width=2.5in]{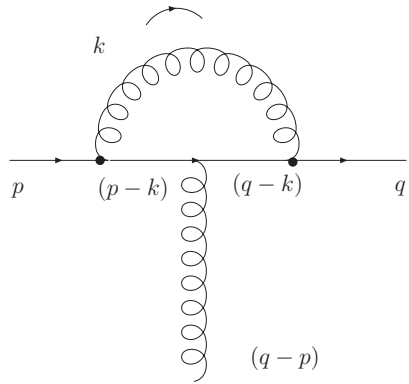}
\end{center}
\caption{The one-loop vertex diagram.} \label{abelianvert}
\end{figure}
The amplitude for the vertex diagram, Fig. (\ref{abelianvert}), 
in the commutative setting was calculated in Ref. 
\cite{Chaichian2}. The corresponding vertex amplitude in the NC 
case is given by
    \be
\GG_\m= - \frac{e^2}{M} e^{\frac{i}{2}p\times q} \intk 
\e_{\l\s\r}\frac{k^{\r}}{k^2}{\bar 
u}(q)\frac{\g^{\s}(\slash{q}+\slash{k}+m)\g_{\m} 
(\slash{p}+\slash{k}-m)\g^{\l}} {[(q+k)^2-m^2][(p+k)^2-m^2]}u(p) 
e^{i k \times \CK}.
    \ee
In the above $\CK_\m = (q - p)_\m$. Here the gamma matrices are 
defined as $\g^0=\s_2, \g^1=i\s_3, \g^3=i\s_1$, and $g_{\m\n}$ is 
taken to be diag(1,-1,-1) same as that adopted in Ref. 
\cite{Deser}. The above integral can be simplified using the 
identity $\g_\m\g_\n=g_{\m\n}-i\e_{\m\n\r}\g^{\r}$ and the 
mass-shell conditions: $(\slash{p}-m)u(p)=0$ and 
$\bar{u}(q)(\slash{q}-m)=0$. After some lengthy algebra and 
defining $\CP_\m = (q+p)_\m$ the above amplitude can be cast in 
the form
    \bea \no
\no \GG_\m & = & - \frac{e^2}{M} e^{\frac{i}{2}p\times q} \intk 
\left[\frac{4i(m\g_{\m}-q_{\m}-{\cal P}_{\m})}{(2q \cdot k + k^2) 
(2p \cdot k + k^2)}+ \frac{i\g_{\m}\slash{k}}{k^2(2p\cdot k+k^2)} 
+ \frac{i\slash{k}\g_{\m}}{k^2(2 q \cdot k + k^2)} \right.
\\ & + & \left. \frac{4\e_{\l\s\r}k^{\r}q^{\s}p^{\l}\g_{\m}}
{k^2(2q \cdot k + k^2) (2p \cdot k + k^2)} \right] e^{i k \times 
\CK}.
    \eea
Terms proportional to $k_\m$ which were zero in the commutative 
case, due to symmetry arguments, cannot be dropped because of a 
momentum dependent phase $\exp(ik\times \CK)$. Writing the above 
integral as $\GG_\m=\GG_{\m}^{(1)}+ \GG_{\m}^{(2)}+ 
\GG_{\m}^{(3)}+ \GG_{\m}^{(4)}$ for the sake of convenience, and 
performing the standard noncommutative loop integrations yield
    \bea \no
\GG_{\m}^{(1)} = -\frac{4ie^2}{M(\sqrt{\pi^3})^3}\, 
e^{-\frac{i}{2}p\times q}\int _{0}^{1}dx \left\{ 2i(m\g_{\m}+x 
\CK_{\m}+p_{\m}- \CP_{\m}) 
\left[\frac{|\tilde{\CK}|}{2|\D_1|}\right]^{1/2} 
K_{1/2}(|\tilde{\CK}| |\D_1|)
\right.\\
\left. + \tilde{\CK}_\m 
\left[\frac{2|\D_1|}{|\tilde{\CK}|}\right]^{1/2} 
K_{-1/2}(|\tilde{\CK}| |\D_1|) \right\},
    \eea
    \bea \nonumber
\GG_{\m}^{(2)} = \frac{i e^2 \g_\m}{M (2\sqrt{\pi^3})^3}\, 
e^{\frac{i}{2}p\times q}\int _{0}^{1} dx \, e^{-ixp \times q} 
\left\{\slash{\tilde{\CK}} \left[\frac{2|\D_2|}
{|\tilde{\CK}|}\right]^{1/2} K_{-1/2}(|\tilde{\CK}| |\D_2|) \right.\\
\left. + 2 i x \slash{p} 
\left[\frac{|\tilde{\CK}|}{2|\D_2|}\right]^{1/2}K_{1/2}(|\tilde{\CK}| 
|\D_2|) \right\},
 \eea
\bea \nonumber \GG_{\m}^{(3)} = \frac{ie^2}{M(2\sqrt{\pi^3})^3}\, 
e^{\frac{i}{2}p\times q} \int_{0}^{1} dx e^{-i x p \times q} 
\left\{\slash{\tilde{\CK}} \left[\frac{2|\D_3|} 
{|\tilde{\CK}|}\right]^{1/2} K_{-1/2}(|\tilde{\CK}| |\D_3|)
\right.\\
\left. + 2 i x \slash{q}
\left[\frac{|\tilde{\CK}|}{2|\D_3|}\right]^{1/2} 
K_{1/2}(|\tilde{\CK}| |\D_3|) \right\}\g_\m, \eea
    \be
\GG_{\m}^{(4)} = - \frac{e^2 \e_{\l \s \r} q^\s p^\l \g_\m}{M (2
\sqrt{\pi^3})^3} \, e^{\frac{i}{2}p \times q} \int_{0}^{1} dx 
\int_{0}^{x} dy \, e^{-i(1-y) p \times q}\,  
\tilde{\CK}^{\r}\left[\frac{|\tilde{\CK}|}{2 |\D_4|}\right]^{1/2} 
K_{1/2}(|\tilde{\CK}||\D_4|).
    \ee
Here and in what follows, a tilde over a momentum indicates that 
it is contacted with the NC parameter $\th$ \textit{i.e.}, 
$\tilde{\CK}^\m \equiv \th^{\m \n} \CK_\n$. Furthermore we have 
abbreviated $\D_1^2=(x \CK + p)^2$, $\D_2^2=m^2 x^2$, $\D_3^2= 
m^2 x^2$ and $\D_4^2=m^2(1-y)^2-(x-y)(1-x) \CK^2$. In obtaining 
the above integrals we have used
    \be
\int_{0}^{\infty}dx \, x^{\n-1} e^{-\g x - \b/x} = 2 
(\b/\g)^{\n/2} K_{\n}[2\sqrt{\b\g}],
 \ee
where $K_{\n}$ is the modified Bessel function of the second 
kind. The parametric integrals can be solved elegantly by going 
over to the rest frame of the electron which implies $p \times q 
=0$. Using
    \be
K_{\pm 1/2}(z) = \sqrt{\frac{\pi}{2}} \frac{e^{-z}}{\sqrt{z}}
    \ee 
and retaining terms to only first order in $\th$ we get
    \bea \no
\GG_{\m}^{(1)} &=& -\frac{ie^2}{2\pi M} \int_{0}^{1}dx \left\{i[m 
\g_\m + x \CK_\m + p_\m - 
\CP_\m]\left[\frac{1}{|\D_1|}-|{\tilde{\CK}}| \right] +
{\tilde{\CK}}_\m \left[\frac{1}{|{\tilde{\CK}}|}- |\D_1| \right]\right\} \\
\no \GG_{\m}^{(2)} &=& \frac{ie^2\g_\m}{8 \pi M} \int_{0}^{1}dx 
\left\{\slash{\tilde{\CK}}\left[\frac{1}{|{\tilde{\CK}}|}- 
|\D_2|\right] + i \, x \, \slash{p} \left[
\frac{1}{|\D_2|}-|{\tilde{\CK}}| \right] \right\}
\\ \no \GG_{\m}^{(3)} &=& \frac{ie^2}{8 \pi M} \int_{0}^{1}dx
\left\{\slash{\tilde{\CK}} \left[\frac{1}{|{\tilde{\CK}}|}-|\D_3| 
\right] + i\, x \, \slash{q} \left[\frac{1}{|\D_3|}-|{\tilde{\CK}}|\right]
\right\} \g_{\m} \\
\GG_{\m}^{(4)} &=& - \frac{e^2 \e_{\l\s\r} q^\s p^\l 
{\tilde{\CK}}^\r \g_\m }{4 \pi M}\int_{0}^{1}dx \int_{0}^{x} dy 
\left\{\frac{1}{|\D_4|} - \frac{|\tilde{\CK}|}{2} \right\}.
 \eea
Solving the integrals in the low momentum transfer limit 
\textit{i.e.}, $\CK^2=0$ and making use of the three dimensional 
analogue of Gordon's decomposition
    \be
\g_\m=\frac{1}{2m}[\CP_\m + i \e_{\m\n\l} \CK^\n \g^\l ],
    \ee
the various contributions to the one-loop vertex can be written as
    \bea \no
\GG^{(1)}_\m & = & \frac{ie^2}{4\pi M}\left[ \frac{\e_{\m\n\l} 
\CK^{\n} \g^{\l}}{m} - \e_{\m\n\l} \CK^{\n} \g^{\l} |\tilde{\CK}| 
- 2 \tilde{\CK}_\m \left( \frac{1}{|\tilde{\CK}|} - m \right) 
\right] \\ \no 
 \GG^{(2+3)}_\m & = & \frac{ie^2}{4\pi M}\left[ i 
\g_\m - \frac{i\g_\m m |\tilde{\CK}|}{2} + 
 \tilde{\CK}_\m \left(\frac{1}{|\tilde{\CK}|} - \frac{m}{2} \right) 
 \right] \\
\GG^{(4)}_\m & = & - \frac{e^2}{4\pi M} \g_{\m} \e_{\l\s\r} q^\s 
p^\l \tilde{\CK}^{\r} \left[ \frac{1}{m} - 
\frac{|\tilde{\CK}|}{2} \right].
    \eea

\section{Conclusions}

Now that the total contribution for the vertex has been obtained 
let us study the result. The first terms of $\GG^{(1)}$ and 
$\GG^{(2+3)}$ are the usual commutative MM and vertex 
contributions respectively. The second terms of $\GG^{(1)}$ and 
$\GG^{(2+3)}$ give the NC modulated corrections to the MM and the 
vertex. As can be seen from the signs of these terms they are 
opposite to that of the commutative contributions. Hence we 
expect that the second term in $\GG^{(1)}$ will reduce the total 
MM. But then of interest are the third terms of $\GG^{(1)}$ and 
$\GG^{(2+3)}$ which happen to be a pure $\th$ induced 
contributions to the MM. Such terms have been in the case 
ordinary QED as well \cite{Riad}. These terms as can be seen, the 
one with $1/|\tilde{\CK}|$ as the coefficient leads to a 
reduction in the MM whereas that arising from $m$ leads to the 
increase. The final contribution $\GG^{(2+3)}$ the second term is 
higher order in $\th$ and hence need not bee considered here. The 
first term is higher order in the derivatives and will not 
contribute to the MM.

\acknowledgments

I thank A. P. Balachandran and Prasanta K. Panigrahi for many 
useful discussions.

\end{document}